\documentclass[11pt,english,aps,manuscript]{revtex4}

\usepackage[T1]{fontenc}
\usepackage[latin9]{inputenc}
\usepackage{textcomp}
\usepackage{amsmath}

\makeatletter

\providecommand*{\perispomeni}{\char126}
\AtBeginDocument{\DeclareRobustCommand{\greektext}{%
  \fontencoding{LGR}\selectfont\def\encodingdefault{LGR}%
  \renewcommand{\~}{\perispomeni}%
}}
\DeclareRobustCommand{\textgreek}[1]{\leavevmode{\greektext #1}}
\DeclareFontEncoding{LGR}{}{}

\newcommand{\lyxmathsym}[1]{\ifmmode\begingroup\def\b@ld{bold}
  \text{\ifx\math@version\b@ld\bfseries\fi#1}\endgroup\else#1\fi}

\@ifundefined{textcolor}{}
{%
 \definecolor{BLACK}{gray}{0}
 \definecolor{WHITE}{gray}{1}
 \definecolor{RED}{rgb}{1,0,0}
 \definecolor{GREEN}{rgb}{0,1,0}
 \definecolor{BLUE}{rgb}{0,0,1}
 \definecolor{CYAN}{cmyk}{1,0,0,0}
 \definecolor{MAGENTA}{cmyk}{0,1,0,0}
 \definecolor{YELLOW}{cmyk}{0,0,1,0}
 }

\makeatother

\usepackage{babel}

\begin{document}

\title{Chern-Simons Gravity with (Curvature)$^{2}$- and (Torsion)$^{2}$-Terms
and A Basis of Degree-of-Freedom Projection Operators}

\author{J. A. Helayël-Neto}

\email{helayel@cbpf.br}

\affiliation{Centro Brasileiro de Pesquisas Físicas, Rua Dr. Xavier Sigaud 150,
Urca,\\
 Rio de Janeiro, Brazil, CEP 22290-180}

\author{C. A. Hernaski}

\email{carlos@cbpf.br}

\affiliation{Centro Brasileiro de Pesquisas Físicas, Rua Dr. Xavier Sigaud 150,
Urca,\\
 Rio de Janeiro, Brazil, CEP 22290-180}

\author{B. Pereira-Dias}

\email{bpdias@cbpf.br}

\affiliation{Centro Brasileiro de Pesquisas Físicas, Rua Dr. Xavier Sigaud 150,
Urca,\\
 Rio de Janeiro, Brazil, CEP 22290-180}

\author{A. A. Vargas-Paredes}

\email{alfredov@cbpf.br}

\affiliation{Centro Brasileiro de Pesquisas Físicas, Rua Dr. Xavier Sigaud 150,
Urca,\\
 Rio de Janeiro, Brazil, CEP 22290-180}

\author{V. J. Vasquez-Otoya}

\email{vjose@cbpf.br}

\affiliation{Centro Brasileiro de Pesquisas Físicas, Rua Dr. Xavier Sigaud 150,
Urca,\\
 Rio de Janeiro, Brazil, CEP 22290-180}

\pacs{01.2k?}
\begin{abstract}
We investigate the effects of (Curvature)$^{2}$- and (Torsion)$^{2}$-
terms in the Einstein-Hilbert-Chern-Simons Lagrangian. The purposes
are two-fold: (i) to show the efficacy of an orthogonal basis of degree-of-freedom
projection operators recently proposed and to ascertain its adequacy
for obtaining propagators of general parity-breaking gravity models
in three dimensions; (ii) to analyze the role of the topological Chern-Simons
term for the unitarity and the particle spectrum of the model squared-curvature
terms in connection with dynamical torsion. Our conclusion is that
the Chern-Simons term does not influence the unitarity conditions
imposed on the parameters of the Lagrangian, but significantly modifies
the particle spectrum.
\end{abstract}
\maketitle

\section{Introduction}

In connection with the AdS/CFT correspondence in three dimensions,
planar quantum gravity has been the object of renewed and raising
interest\cite{Carlip:1995zj}. As showed in Ref. \cite{btz}, Einstein-Hilbert
(E-H) gravity with a negative cosmological constant has black hole
solutions and for this reason it is interestingly related to two dimensional
CFT theories on the AdS boundary. In the work of \cite{Witten:2007kt},
Witten has reassessed other relevant aspects of three-dimensional
gravity. In a subsequent paper, Li, Song and Strominger \cite{Li:2008dq}
have proposed a chiral gravity model in three space-time dimensions
and they focus their efforts to study topologically massive gravity.

In spite of E-H gravity in three dimensions having no propagating
degrees of freedom, the introduction of (curvature)$^{2}$- terms
allow new propagating modes. However, unitarity could be jeopardized
due to the higher-derivative terms. Surprisingly, a recently proposed
higher-derivative model, known as Bergshoeff-Hohm-Townsend model \cite{Bergshoeff:2009aq,Bergshoeff:2009hq},
was shown to be unitary and renormalizable \cite{Nakasone:2009bn,Oda:2009ys}.
This model is a specific combination of the E-H action with the wrong
sign added with higher-derivative curvature terms, which is equivalent
to the Pauli-Fierz Lagrangian at the linearized level.

In odd dimensional theories, it is tempting to consider a Chern-Simons
term. In the case of vector fields, this term gives to the photon
a mass in a gauge invariant way. For planar gravity, this term was
first considered by Deser, Jackiw and Templeton in Ref. \cite{deser-jackiw-templeton}.
In fact, the E-H Lagrangian with the wrong sign added to the Chern-Simons
term propagates a parity-breaking massive spin-$2$ mode. In spite
of the fact that gravitational Chern-Simons term has three derivatives,
it was shown that this model has neither ghosts nor acausalities.
Furthermore, the presence of the three derivatives suggests that the
ultraviolet divergences of the model could be stabilized rendering
it power-counting renormalizable. Actually, this was explicitly shown
in Ref. \cite{Deser:1990bj,Keszthelyi:1991ha,Oda:2009im}. The issue
of unitarity in extensions of these theories, such as the incorporation
of quadratic terms, is not a trivial matter, as discussed in \cite{Accioly:2001,Accioly:2003hg,Accioly:2004qw,Dalmazi:2009bd}.

In four dimensions, massive gravity theories are motivated by the
outstanding result that they could suitably modify Einstein's General
Relativity at very large distance scales (actually, cosmological scales),
in such a way that the present accelerated expansion of our universe
may be taken into account without invoking the idea of dark energy.
Gravity models with dynamical torsion may naturally yield a way to
generate mass for gravitons. So, we take the viewpoint that, by understanding
torsion propagation and massive gravitons in three-dimensional space,
we may get insights that can be helpful for a better comprehension
of massive gravitation in connection with dynamical torsion in four
dimensions.

Once we are convinced of the relevance of investigating aspects of
quantum gravity in three-dimensional space-time, we could go further
and try to understand how the degrees of freedom associated to torsion
may influence and affect properties of planar quantum gravity, previously
contemplated in the absence of torsion. Our work sets out just to
pursue an investigation of the possible effects torsion may induce
on quantum-mechanical aspects of planar gravity. Of special interest
for us is the emergence of massive gravitons, once torsion is allowed
to be dynamical. For a deeper discussion on the role of torsion in
four-dimensional quantum gravity, we address the reader to the works
of Ref. \cite{Shapiro:1998zh,Cho:2009fa,Kim:2006yv,Hammond:2002rm,necessitytorsion}.

In this paper, we shall be mainly interested in Chern-Simons topological
term, but with the modification that shall allow propagating torsion.
This is appealing whenever we adopt the vielbein formalism for gravitation,
since, as shown in \cite{Witten:1988hc,Cacciatori:2005wz}, the spin
connection and vielbein can be combined in such way that they form
a connection for a truly gauge theory. In this way, the field-strength
related to this connection carries both the Riemann and torsion tensor.
Our main goal is to investigate the spectrum and unitary properties
of the topological gauge gravity theory in the first-order formalism.
By this, we mean the E-H Lagrangian with the Chern-Simons term considering
the spin connection and vielbein as independent fields. Besides that,
we also consider (curvature)$^{2}$- and (torsion)$^{2}$- terms,
which renders our analysis more general.

In order to analyze the spectral properties of the model, the attainment
of the propagator becomes a primary goal. There are various methods
for the deriving propagators. The algebraic method based on the spin
projection operators has been shown very efficient and has been widely
used in the literature \cite{neville,S-N,our}. A three-dimensional
analogous basis to the one proposed in \cite{S-N} is not enough to
handle with the problem proposed in this paper. One possibility to
circumvent this issue is to extend the basis by means of the decomposition
of the transverse operator, $\theta$, and the longitudinal operator,
$\omega$, into more fundamental ones. In three dimensions, the procedure
is to decompose $\theta$, which is a definite spin projector, into
two degree-of-freedom projectors, $\rho$ and $\sigma$. In \cite{extending the spin projection operators},
the set of operators in terms of $\left\{ \theta,\ \omega\right\} $
is rewritten in terms of $\rho$, $\sigma$ and $\omega$. As it becomes
clear in that work, additional mapping operators among the projectors
are needed, since we are dealing with degree-of-freedom projectors
rather than spin projectors. The physical appeal for this decomposition
has to do with the different role played by parity in four and three
dimensions. This extension of the basis enables one to handle terms
with an explicit Levi-Civita tensor, which is necessary to describe
parity-breaking and to take advantage of the dual aspects of the fields.
The propagator and the conditions for the absence of ghosts and tachyons
are, by this procedure, directly obtained. 

Our results allow us to discuss the role played by the Chern-Simons
term on the conditions for the unitarity and in the spectrum properties
of the model. We also explicitly analyze the case where we discard
the Chern-Simons term and compare them with the results obtained in
\cite{our} for D=3, in order to verify the consistency of the proposed
basis. 

The outline of this paper is as follows: in Sec. II, we present the
model and the conventions used in this work. In Sec. III, we work
out the propagator of the model. Sec. IV tackles the issue of spectrum
consistency by analyzing the conditions for absence of ghosts and
tachyons in the massive and massless sectors. Finally, in Sec. V,
we set up our Concluding Remarks. An Appendix follows, where we list
the inverse of the spin matrices used to calculate the propagators
of Sec. III.

\section{Description of The Model\label{sec:Description-of-The}}

For the sake of generality, we investigate a general gravity Lagrangian
which include quadratic terms in the curvature and torsion added to
a Chern-Simons term:\begin{eqnarray}
\mathcal{L} & = & e[\left(u-s\right)R+\frac{1}{8}\left(3\left(\beta+\gamma\right)-\frac{1}{4}\xi\right)R^{2}+\beta R_{\mu\nu}R^{\mu\nu}+\gamma R_{\mu\nu}R^{\nu\mu}+\label{eq:lagrangian}\\
 &  & \frac{1}{8}\left(u+r+2s\right)T_{abc}T^{abc}+\frac{1}{4}\left(u+r-2s\right)T_{abc}T^{bca}+\frac{1}{2}\left(u-2s-t\right)T_{ab}^{\ \ b}T_{\ \ c}^{ac}]+d\mathcal{L}_{CS},\nonumber \end{eqnarray}
where $r$, $s$, $t$, $u$, are arbitrary parameters with mass dimension
equal to $1$, whereas $\xi$, $\beta$ and $\gamma$ are inverse
mass parameters and $d$ is dimensionless. Also, we should justify
that this apparently unusual combination of parameters is a mere matter
of convenience. Actually, in our analysis of the spectral conditions
(subject of Section \ref{sec:Spectral-Consistency-Analysis}), the
parameters associated to the terms of Lagrangian density above combine
in such a way that the form we propose in \ref{eq:lagrangian} yield
to considerable algebraic simplifications, without any loss of generality
in our results. The term \begin{equation}
\mathcal{L}_{CS}=\epsilon^{\alpha\beta\gamma}(R_{\alpha\beta ab}\omega_{\gamma}^{\ ab}+\frac{2}{3}\omega_{\alpha b}^{\ \ c}\omega_{\beta c}^{\ \ a}\omega_{\gamma a}^{\ \ b})\end{equation}
is the well-known topological Chern-Simons term. Some remarks are
in order. First, the absence of terms of Riemann squared is due to
the fact that, in three dimensions, the Riemann tensor can be written
in terms of the Ricci tensor and the scalar curvature. Second, the
absence of the cosmological constant and the translational Chern-Simons
term, $\epsilon^{\mu\nu\lambda}T_{\mu\nu}^{\ \ a}e_{\lambda a}$ \cite{Mielke:1991nn},
are due to a more subtle reason. We shall adopt the position of expanding
the graviton field around a Minkowski space. But, it is well-known
that free-matter solutions for theories with these two terms are non-flat.
As an immediate consequence, the introduction of these terms would
spoil the gauge symmetry present in the linearized version of the
model. Third, powers of curvature and torsion higher than two have
not been considered in order to avoid higher derivatives that usually
are hazardous to unitarity properties of the model. Finally, since
we are considering the Chern-Simons term and thus a parity-broken
model, there is the possibility of considering quadratic terms built
up from the dual of torsion and Ricci tensor, as for example, $\epsilon_{\mu}^{\ \kappa\lambda}T_{\kappa\lambda}^{\ \ a}R_{\ a}^{\mu}$,
$R\epsilon^{\mu\nu\kappa}T_{\mu\nu\kappa}$, $\epsilon_{\mu\nu}^{\ \ \kappa}T_{\kappa a}^{\ \ a}R^{\mu\nu}$.
All these mixings couple the vielbein and spin-connection field strengths
in a non-trivial way.  It is interesting to notice that these terms
may be regarded as originating from a Lorentz-symmetry violating gravity
model in ($1+3$)-dimensions in presence of a time-like background
vector, $v^{\mu}$. The terms: $\epsilon^{\mu\nu\kappa\lambda}v_{\mu}T_{\nu\kappa}^{\ \ a}R_{\lambda a}$,
$\epsilon^{\mu\nu\kappa\lambda}v_{\mu}RT_{\nu\kappa\lambda}$ and
$\epsilon^{\mu\nu\kappa\lambda}v_{\mu}T_{\nu a}^{\ \ a}R_{\kappa\lambda}$
yield, respectively, the $3D$-terms mentioned above for a time-like
background vector, $v^{\mu}=(\mu;\vec{0})$. In connection with a
paper by Kostelecký \cite{Kostelecky:2003fs} and the work of Ref.
\cite{boldo}, an investigation of the possible origin and consequences
of such $4D$-terms in the spectrum of gravity models with deviations
from Lorentz symmetry and a full study of $3D$-gravity with the inclusion
of the T-R-type terms above demands special attention and is the subject
an investigation we are pursuing. 

In this paper, we shall work in the first-order formalism, where the
vielbein ($e_{\mu}^{a}$) and spin connection ($\omega_{\mu}^{\ ab}$
) are taken as fundamental fields. We also set up the conventions
for the Minkowski metric, $\eta_{\mu\nu}=\left(+1,-1,-1\right),$
the Levi-Civita symbol, $\epsilon_{012}=+1$, and the Riemann and
torsion tensors,\begin{subequations}\begin{eqnarray}
 & R_{\mu\nu}^{\ \ ab} & =\partial_{\mu}\omega_{\nu}^{\ ab}-\partial_{\nu}\omega_{\mu}^{\ ab}+\omega_{\mu\ c}^{\ a}\omega_{\nu}^{\ cb}-\omega_{\nu\ c}^{\ a}\omega_{\mu}^{\ cb},\label{eq:curvature}\\
 & T_{\mu\nu}^{\ \ a} & =\partial_{\mu}e_{\nu}^{\ a}-\partial_{\nu}e_{\text{\ensuremath{\mu}}}^{\ a}+\omega_{\mu\ c}^{\ a}e_{\nu}^{\ c}-\omega_{\nu\ c}^{\ a}e_{\mu}^{\ c},\label{eq:torsion}\end{eqnarray}
\end{subequations}where the Greek indices refer to the world manifold
and Latin ones for the frame indices. The contracted tensors are\begin{subequations}\begin{eqnarray}
 & R_{\mu}^{\ a} & =e_{\ b}^{\nu}R_{\mu\nu}^{\ \ ab},\\
 & R & =e_{\ a}^{\mu}e_{\ b}^{\nu}R_{\mu\nu}^{\ \ ab}.\end{eqnarray}
\end{subequations}In order to settle down a quantum theory, we shall
consider the following fluctuation around the Minkowski vacuum:\begin{subequations}\begin{eqnarray}
 & e_{\mu}^{\ a} & =\delta_{\mu}^{\ a}+\tilde{e}_{\mu}^{\ a}\\
 & \omega_{\mu}^{\ ab} & =\tilde{\omega}_{\mu}^{\ ab}\end{eqnarray}
\end{subequations}Henceforth, the distinction between Greek and Latin
indices becomes unnecessary. It is also convenient to decompose the
vielbein fluctuation, $\tilde{e}_{ab}$, into its symmetric, $\phi_{ab}$,
and antisymmetric, $\epsilon_{ab}^{\ \ c}\chi_{c}$, components. An
analogous decomposition is done for the dual field of the spin connection
fluctuation, being $\psi_{ad}$ the symmetric and $\epsilon_{ad}^{\ \ e}\lambda_{e}$
the antisymmetric components:\begin{subequations}\begin{eqnarray}
 & \tilde{e}_{ab} & =\phi_{ab}+\epsilon_{ab}^{\ \ c}\chi_{c},\label{eq:vielbein in terms of duals}\\
 & \tilde{\omega}_{abc} & =\epsilon_{bc}^{\ \ d}\left(\psi_{ad}+\epsilon_{ad}^{\ \ e}\lambda_{e}\right).\label{eq:spin connection in terms of duals}\end{eqnarray}
\end{subequations}In the sequel, we shall consider $\phi$, $\chi$,
$\psi$ and $\lambda$ as the fundamental fields of the linearized
model.

\section{Calculation of The Propagator\label{sec:Calculation-of-The}}

It is our main goal to analyze the spectral consistency of the model
\eqref{eq:lagrangian}. These aspects can be readily obtained by means
of the propagator of the model. In order to accomplish this task,
let us consider the Lagrangian up to second-order terms in the quantum
fluctuations,\begin{equation}
\left(\mathcal{L}\right)_{2}=\frac{1}{2}\sum_{\alpha,\beta}\Psi_{\alpha}\mathcal{O_{\alpha\beta}}\Psi_{\beta},\label{eq:linear lagrangian}\end{equation}
where $\Psi_{\alpha}$ is a multiplet that carries the 18 components
$\left(\phi_{ab},\psi_{ab},\chi_{a},\lambda_{a}\right)$ and $\mathcal{O}_{\alpha\beta}$
is the wave operator which contains $\eta$'s, $\epsilon$'s and at
most two derivatives. The saturated propagator is written as\begin{equation}
\Pi=i\sum_{\alpha,\beta}\mathcal{S}_{\alpha}^{*}\mathcal{O}_{\alpha\beta}^{-1}\mathcal{S}_{\beta},\label{eq:propagator}\end{equation}
with $\mathcal{S}_{\alpha}$ being the sources for the fundamental
fields.

The problem of the attainment of the propagator is reduced to the
problem of inversion of the wave operator. Once one has a complete
basis in which the wave operator can be expanded, to invert the operator
becomes a lengthy, but straightforward task.

A first step to treat the attainment of the propagator for Chern-Simons
gravity in second-order formalism was carried out in \cite{Pinheiro:1997ka,Pinheiro:1997kc},
with the help of an extension of the Barnes-Rivers operators\cite{Pinheiro:1992wv}.
However, the consideration of Lagrangians with a larger number of
free parameters, specially in first-order formalism, makes this task
technically difficult \cite{comments on tmg,discussing}, due to the
non-trivial algebra that these operators satisfy.

A more efficient technique to simplify this issue is to decompose
the wave operator of the linearized Lagrangian in an orthogonal basis
of projector operators. In \cite{neville,S-N}, an orthonormal basis
of spin-parity operators in $4D$ is proposed which is suitable to
study spectral properties of parity-preserving models containing a
rank-2 tensor and a rank-3 tensor anti-symmetric in two indices. The
advantage of an orthonormal basis of spin-parity operators is the
decomposition of the wave operator into spin-parity sectors that can
be inverted independently. Gravity Lagrangians in first-order formalism
are fairly-well accommodated in this treatment. Also, the gauge symmetries
of the model are conveniently handled. Its generalization for arbitrary
space-time dimensions \cite{our} is straightforward and the particular
three-dimensional case is the one relevant to this paper.

It is known that the spin-parity operator basis obtained as a generalization
from the one proposed in \cite{S-N} cannot handle the Chern-Simons
term in a straightforward way, since the wave operator contains the
Levi-Civita symbol. In others words, this set of operators does not
form a basis for parity-breaking Lagrangians, which is the sort of
model where the Chern-Simons term is encoded in. In order to circumvent
this problem, it is proposed in \cite{extending the spin projection operators}
a parity operator basis in 3-D that makes possible to analyze parity-breaking
gravity models on the same foot as those that do not violate parity.
In \cite{extending the spin projection operators}, it is argued that
each spin decomposition of the field does not have a definite parity
and finding out degree-of-freedom operators is the convenient way
to set up a basis operator in this case. With the mentioned basis
at hand, we perform the wave operator decomposition of the linearized
Lagrangian. Further discussions and a list of these operators are
given in \cite{extending the spin projection operators}.

With such a basis, one expands the wave operator as

\begin{equation}
\mathcal{O}_{\alpha\beta}=\sum_{J,ij}a_{ij}^{\varphi\vartheta}\left(J\right)P_{ij}^{\varphi\vartheta}\left(J^{PQ}\right)_{\alpha\beta}.\end{equation}
Let us clarify the notation. The diagonal operators $P_{ii}^{\varphi\varphi}\left(J^{PP}\right)$
are projectors in each of the degrees of freedom of the spin $\left(J\right)$
and parity $\left(P\right)$ sectors of the field $\varphi$, while
the $P_{ij}^{\varphi\vartheta}(J^{PQ})$ (with $i\neq j$) are mappings
between the projectors $P_{ii}^{\varphi\varphi}\left(J^{PP}\right)$
and $P_{jj}^{\vartheta\vartheta}\left(J^{QQ}\right)$. This can be
read off in the following relations:

\begin{equation}
\sum_{\beta}P_{ij}^{\Sigma\Psi}\left(I^{PQ}\right)_{\alpha\beta}P_{kl}^{\Lambda\Xi}\left(J^{RS}\right)_{\beta\gamma}=\delta_{jk}\delta^{\Psi\Lambda}\delta^{IJ}\delta^{QR}P_{il}^{\Sigma\Xi}\left(I^{PS}\right)_{\alpha\gamma},\end{equation}
\begin{equation}
\sum_{i,J^{PP}}P_{ii}\left(J^{PP}\right)_{\alpha\beta}=\delta_{\alpha\beta}.\end{equation}

The $a_{ij}^{\Sigma\Lambda}\left(J\right)$ are the coefficient in
the wave operator expansion. These can be arranged in matrices representing
the contribution to the spin $\left(J\right)$. When these matrices
are non-singular, the saturated propagator \eqref{eq:propagator}
is given by:

\begin{equation}
\Pi=i\sum_{\alpha,\beta,J^{PQ}}\mathcal{S}_{\alpha}^{*}a_{ij}^{-1\varphi\vartheta}\left(J\right)P_{ij}^{\varphi\vartheta}\left(J^{PQ}\right)_{\alpha\beta}\mathcal{S}_{\beta.}\end{equation}

However, the considered Lagrangian \eqref{eq:lagrangian} is invariant
under local Lorentz and general coordinate transformations. This implies
that the linearized Lagrangian is invariant under some local transformations
of the fields. Gauge invariance makes the coefficient matrices to
become degenerate. In Ref. \cite{neville}, it is shown that the correct
gauge invariant propagator is obtained by taking the inverse any largest
non-degenerate sub-matrix and then saturating it with sources.

For the model \eqref{eq:lagrangian}, the coefficients $a_{ij}^{\Sigma\Lambda}\left(J\right)$
form the $6\times6$ spin-0, $8\times8$ spin-1 and $4\times4$ spin-2
matrices. The spin-0 and spin-1 matrices are degenerate. We list below
the largest non-degenerate sub-matrices obtained from them:\begin{subequations}

\begin{equation}
a\left(0\right)=\left(\begin{array}{cccc}
2u+4r+2\left(\beta-\gamma\right)p^{2} & 2\sqrt{2}r & 0 & 8\sqrt{2}id\sqrt{p^{2}}\\
2\sqrt{2}r & 2\left(u+r\right) & 0 & 0\\
0 & 0 & 2\left(u-t-s\right)p^{2} & 2\sqrt{2}i\sqrt{p^{2}}t\\
-8i\sqrt{2}d\sqrt{p^{2}} & 0 & -2\sqrt{2}it\sqrt{p^{2}} & -4t+\xi p^{2}\end{array}\right),\label{eq:spin0}\end{equation}

\begin{equation}
a\left(1\right)=\left(\begin{array}{cccc}
2u+\beta p^{2} & -4id\sqrt{p^{2}} & 0 & -iu\sqrt{p^{2}}\\
4id\sqrt{p^{2}} & 2u+\beta p^{2} & iu\sqrt{p^{2}} & 0\\
0 & -iu\sqrt{p^{2}} & \frac{1}{2}\left(u-t\right)p^{2} & 0\\
iu\sqrt{p^{2}} & 0 & 0 & \frac{1}{2}\left(u-t\right)p^{2}\end{array}\right),\label{eq:spin1}\end{equation}
\begin{equation}
a\left(2\right)=\left(\begin{array}{cccc}
2u+2\left(\beta+\gamma\right)p^{2} & 8id\sqrt{p^{2}} & 0 & 2iu\sqrt{p^{2}}\\
-8id\sqrt{p^{2}} & 2u+2\left(\beta+\gamma\right)p^{2} & -2iu\sqrt{p^{2}} & 0\\
0 & 2iu\sqrt{p^{2}} & 2sp^{2} & 0\\
-2iu\sqrt{p^{2}} & 0 & 0 & 2sp^{2}\end{array}\right),\label{eq:spin2}\end{equation}
\end{subequations} where $p^{2}=p_{a}p^{a}$, with $p^{a}$ being
the relativistic three-momentum. Their inverses, needed for the attainment
of the propagators, are given in the Appendix.

In Ref. \cite{our}, one considers the same Lagrangian \eqref{eq:lagrangian},
except for the Chern-Simons term, in an arbitrary space-time dimension.
So, it is worthwhile to compare our results so far, whenever $d=0$,
with those in \cite{our}, for $2+1$ dimensions ($D=3$), in order
to verify the consistency of new basis of operators. At a first glance,
one can notice that there are three more matrices than in our treatment.
In fact, the spin-$2^{-}$and spin-$0^{-}$, which are contained in
the spin-connection field decomposition, cannot appear here since
the spin operators associated with these spins are identically zero
in three dimensions. It also can be verified that the spin operators
associated with the spin-$1^{+}$, in that work, are mapped into spin-$0$
operators when we use the duality relations for the fields. This is
noticed in the spin-$0$ matrix above: for $d=0$, it becomes block-diagonal
with the blocks corresponding to the spin-$0^{+}$ and spin-$1^{+}$
that appear in \cite{our}. The spin-$2$ and spin-$1$ matrices,
compared with the spin-$2^{+}$ and spin-$1^{-}$, remain essentially
the same. The differences are some rearrangements in the parameters
of spin-$1$ and the duplication of the dimension of the matrices
in this work, due to splitting into degrees of freedom instead of
spins. It must be stressed that, by comparing the parameters in both
works, one has to contemplate the fact that in three dimensions the
Riemann tensor can be expressed in terms of the Ricci tensor and scalar
curvature.

\section{Spectral Consistency Analysis\label{sec:Spectral-Consistency-Analysis}}

In this Section, we analyze the spectral consistency of the model.
With this study, we shall impose conditions on the parameters of Lagrangian
\eqref{eq:lagrangian}, in such a way that it does not propagate unphysical
particles, that is, ghosts and tachyons. For the sake of clarity,
we split the discussions for the cases of massive and massless poles.

\subsection{Massive poles}

In terms of the inverse matrices \eqref{eq:InvSpin0}-\eqref{eq:InvSpin2},
we can write the propagator as:

\begin{equation}
\Pi(J^{P})=i\sum_{ij,\alpha,\beta}A_{ij}^{\Sigma\Lambda}\left(J,m^{2}\right)S_{\alpha}^{*}P_{ij}^{\Sigma\Lambda}\left(J^{PQ}\right)_{\alpha\beta}\mathcal{S}_{\beta}(p^{2}-m^{2})^{-1},\end{equation}
where $A(J,m^{2})$ is the $4\times4$ matrix which is degenerate
at the pole $p^{2}=m^{2}$. 

The condition for absence of ghosts and tachyons are respectively
given by:

\begin{equation}
\Im\text{Res}(\Pi|_{p^{2}=m^{2}})>0,\quad\mbox{and}\quad m^{2}>0.\label{eq:ghost/tachyon conditions for massive poles}\end{equation}

The condition for absence of ghosts for each spin is directly related
to the positivity of the matrices $\left(\sum A_{ij}\left(J,m^{2}\right)P_{ij}\right)_{\alpha\beta}.$
However, it can be shown that these matrices have only one non-vanishing
eigenvalue at the pole, which is equal to the trace of $A(J,m^{2})$.
Also, the operators $P_{ij}$ themselves contribute only with a sign
$(-1)^{N}$ when calculated at the pole, where $N$ is the sum of
the number of $\rho$'s and $\sigma$'s in each part of the projector.
Therefore, the condition for absence of ghosts for each spin is reduced
to:

\begin{equation}
(-1)^{N}\mbox{tr}A(J,m^{2})|_{p^{2}=m^{2}}>0.\label{eq:ghost condition in terms of the trace}\end{equation}

Using the conditions \eqref{eq:ghost/tachyon conditions for massive poles}
and \eqref{eq:ghost condition in terms of the trace} for the matrices
\eqref{eq:InvSpin0}-\eqref{eq:InvSpin2}, we have:

\begin{subequations}\begin{eqnarray}
 &  & \mbox{Spin-}\mathbf{2}:\ us(s-u)<0;\ \left(\beta+\gamma\right)>0;\label{eq:unitarityCondSpin2}\\
 &  & \mbox{Spin-}\mathbf{1}:\ \beta<0;\ ut(u-t)<0;\label{eq:unitarityCondSpin1}\\
 &  & \mbox{Spin-}\mathbf{0}:\ (s+t-u)(s-u)t>0;\ (r+u)u(u+3r)<0;\ \xi>0;\ (\beta+\gamma)>0.\label{eq:unitarityCondSpin0}\end{eqnarray}

\end{subequations}

It is remarkable that the conditions for absence of tachyons and ghosts
are equivalent to the ones obtained in \cite{our} in the three-dimensional
case, even if the Chern-Simons term spoils the direct identification
of the respective spin matrices.  

The roots of the matrices denominators \eqref{eq:den0}, \eqref{eq:den1},
and \eqref{eq:den2}, which are given in the Appendix, give us the
masses of the propagating particles. A careful look at the parameter
combination reveals that only the torsion terms are crucial for obtaining
a massive spectrum (as discussed in \cite{our}). This is a remarkable
difference with the second-order formalism for gravity, where the
Chern-Simons term brings up a massive graviton. However, this is due
to the higher-derivative character of such a theory.

The mass spectrum,\begin{subequations}

\begin{eqnarray}
 & \mathbf{2}:\ m_{\pm}^{2} & =\frac{8d^{2}}{\left(\beta+\gamma\right)^{2}}+\frac{u(u-s)}{s\left(\beta+\gamma\right)}\pm\sqrt{\left(\frac{8d^{2}}{\left(\beta+\gamma\right)^{2}}\right)^{2}+2\frac{u(u-s)}{s\left(\beta+\gamma\right)}\frac{8d^{2}}{\left(\beta+\gamma\right)^{2}}},\label{eq:m2}\\
 & \mathbf{1}:\ m_{\pm}^{2} & =\frac{8d^{2}}{\beta^{2}}+\frac{2ut}{\beta\left(u-t\right)}\pm\sqrt{\left(\frac{8d^{2}}{\beta^{2}}\right)^{2}+2\frac{2ut}{\beta\left(u-t\right)}\frac{8d^{2}}{\beta^{2}}},\label{eq:m1}\\
 & \mathbf{0}:\ m_{\pm}^{2} & =\left(\frac{32d^{2}}{\xi\left(\beta-\gamma\right)}+\frac{2t\left(s-u\right)}{\xi\left(s+t-u\right)}-\frac{u\left(3r+u\right)}{2\left(\beta-\gamma\right)\left(r+u\right)}\right)\label{eq:m0}\\
 &  & \pm\sqrt{\left(\frac{32d^{2}}{\xi\left(\beta-\gamma\right)}+\frac{2t\left(s-u\right)}{\xi\left(s+t-u\right)}-\frac{u\left(3r+u\right)}{2\left(\beta-\gamma\right)\left(r+u\right)}\right)^{2}+4\frac{tu\left(s-u\right)\left(3r+u\right)}{\xi\left(\beta-\gamma\right)\left(r+u\right)\left(s+t-u\right)}},\nonumber \end{eqnarray}
\end{subequations}is significantly changed by the Chern-Simons term.
In the spin-$1$ and spin-$2$ sectors the number of particles changes
from one to two. The influence of the Chern-Simons term in the spin-$0$
sector is restricted to shifting the particle masses. All this happens
due to the parity-breaking property of the Chern-Simons term. In fact,
in three dimensions, every non-vanishing spin massive particle has
two degrees of freedom \cite{Binegar:1981gv}. In a parity-preserving
theory, these two degrees of freedom are not distinguished in any
way and, therefore, they propagate in parity doublets. On the other
hand, in a parity-breaking theory, each degree of freedom may propagate
independently. This becomes explicit when one analyses the role of
the Chern-Simons term for the particle masses \eqref{eq:m2}-\eqref{eq:m0}.

\subsection{Massless poles}

For the calculation of the massless propagators, some subtleties require
extra care. The wave operator, as well its inverse, are Lorentz covariant,
thus they can be expressed in terms of the set of following structures:
\begin{equation}
\omega_{ab}=\frac{p_{a}p_{b}}{p^{2}};\ \theta_{ab}=\eta_{ab}-\omega_{ab};\ \epsilon_{abc};\ p_{a}.\label{eq:buildingblocksteta}\end{equation}
As we have discussed earlier, for the attainment of the propagator,
it is extremely convenient to decompose $\theta$ as $\rho+\sigma$
to build an orthonormal set of parity operators. However, for the
calculation of the residue on the massless pole, the explicit dependence
on $p_{a}$ complicates the identification of the spin projectors.
At this stage, we rewrite the propagator in terms of the set \eqref{eq:buildingblocksteta}.

Furthermore, since the model is gauge invariant, there are constraints
that the sources satisfy. They consistently appear in order to inhibit
the non-physical modes from propagating. The explicit expressions
for these constraints are given in terms of the left null-eigenvectors
of the degenerate coefficient matrices:

\begin{equation}
\sum V_{j}^{(L,n)}(J)P_{kj}(J^{PQ})_{\alpha\beta}\mathcal{S}_{\beta}=0.\end{equation}
This equation implies in the following constraints for the fundamental
sources:

\begin{subequations}\begin{eqnarray}
 &  & p^{a}(S_{ab}+S_{ba}+\epsilon_{bca}\Omega^{c})=0,\label{eq:source constraints}\\
 &  & p^{a}(\Sigma_{ab}+\Sigma_{ba})=0,\label{eq:source constraints b}\end{eqnarray}
\end{subequations}where $S_{ab}$, $\Omega_{c}$ and $\Sigma_{ab}$
are the sources for the fields $\psi$, $\lambda$ and $\phi$ respectively.
To compare with previous results, we express the final answer for
the massless propagator in terms of the source to the spin connection
field. The relation among the fields given in \eqref{eq:vielbein in terms of duals}
and \eqref{eq:spin connection in terms of duals} enables us to write
the fundamental sources as:

\begin{subequations}\begin{eqnarray}
 &  & S_{ab}=\frac{1}{2}(\epsilon_{pqb}\tau_{a}^{\ pq}+\epsilon_{pqa}\tau_{b}^{\ pq}),\label{eq:source relations}\\
 &  & \Omega_{a}=-2\eta_{cd}\tau_{\ \ a}^{cd},\label{eq:source relations b}\end{eqnarray}
\end{subequations}with $\tau_{abc}$ being the source of $\omega_{abc}$. 

Using \eqref{eq:source constraints}, \eqref{eq:source constraints b},
\eqref{eq:source relations}, and \eqref{eq:source relations b},
one can show that \begin{eqnarray}
 &  & \Pi\left(p^{2}=0\right)=\frac{1}{2p\lyxmathsym{\texttwosuperior}(s-u)}(\tau^{ab*}\Sigma^{ab\lyxmathsym{\textasteriskcentered}})\left(\begin{array}{cc}
4 & 2i\\
-2i & 1\end{array}\right)\left[\frac{1}{2}(\eta_{ac}\eta_{bd}+\eta_{bc}\eta_{ad})-(\eta_{ab}\eta_{cd})\right](\tau^{de}\Sigma^{de})\nonumber \\
 &  & -\frac{2id}{p\lyxmathsym{\texttwosuperior}(s-u)\lyxmathsym{\texttwosuperior}}\Sigma^{ab\lyxmathsym{\textasteriskcentered}}\left[\frac{1}{2}(\eta_{ac}\epsilon_{dbe}+\eta_{bd}\epsilon_{cae})p^{e}\right]\Sigma^{cd}\label{eq:massless propagator}\\
 &  & +\text{ terms that do not contribute to residue,}\nonumber \end{eqnarray}
where $\tau_{ab}=p^{c}\tau_{abc}$. This result, for $d=0$, agrees
with the one obtained in \cite{our} in the three-dimensional case.
It is shown there that this part of the propagator vanishes. To explicitly
calculate the other term, let us expand the source $\Sigma_{ab}$
as

\begin{equation}
\Sigma_{ab}=c_{1}p_{a}p_{b}+c_{2}(p_{a}\epsilon_{b}+p_{b}\epsilon_{a})+c_{3}\epsilon_{a}\epsilon_{b},\label{eq:source expansion}\end{equation}
where,\begin{subequations}\begin{eqnarray}
 &  & p_{a}=(p_{0},\vec{p}),\\
 &  & q_{a}=(p_{0},-\vec{p}),\end{eqnarray}
\end{subequations}with\begin{subequations}

\begin{eqnarray}
 &  & p^{2}=q^{2}=0,\\
 &  & p.q=(p_{0})^{2}+(\vec{p})^{2},\\
 &  & p.\epsilon=q.\epsilon=0,\\
 &  & \epsilon^{2}=-1.\end{eqnarray}
\end{subequations}

Expansion \eqref{eq:source expansion} is the most general one for
a symmetric rank-$2$ tensor that satisfies \eqref{eq:source constraints}.
Using it in \eqref{eq:massless propagator}, one can show that the
term due to Chern-Simons also identically vanishes. Therefore, even
for $d\neq0$, there is no massless particles propagating in the model
and the relations \eqref{eq:unitarityCondSpin2}-\eqref{eq:unitarityCondSpin0}
are the only ones that must be imposed to ensure the unitarity of
the Lagrangian \eqref{eq:lagrangian}.

\section{Concluding Remarks}

We have considered a general gravity Lagrangian without higher derivatives
and with a parity-breaking Chern-Simons term in the first-order formalism.
It was our interest to investigate the possible unitary Lagrangians
that one can obtain from it and determine the influence of the Chern-Simons
term in the particle spectrum. 

The proposal of applying the basis of spin projection operators developed
in \cite{extending the spin projection operators} was sucessfully
implemented. Two striking features must be emphasized: first, its
orthogonality properties makes the inversion of the wave operators
easier and, second, the analysis of the symmetries of the model becomes
a systematic procedure. Also, the results that appear in the literature
for the same Lagrangian, but without the Chern-Simons term, are recovered,
as it was expected. It became clear that the basis of operator is
not a mere algeabraic convenience. The role of parity in three dimensions
gives the operators a physical meaning. An analogous construction
migth be implemented for dealing with Lorentz-breaking theories in
four dimensions, since parity-breaking terms in three dimensions are
intrinsically related to Lorentz-symmetry-breaking terms in four dimensions.

From the analysis of the spectral consistency (Section \ref{sec:Spectral-Consistency-Analysis}),
we see that the Chern-Simons term does not modify the unitary relations.
Therefore, the possible unitary Lagrangians are the same as those
obtained in \cite{our} added up with the Chern-Simons term. The main
contribution of this term, due to its parity-breaking character, is
to raise the number of massive particle in the spin-$1$ and spin-$2$
sectors, as well as shifting their masses. Furthermore, only massive
modes propagate. In the particular case of Einstein-Hilbert-Chern-Simons
Lagrangian in the first-order formalism, there are no massive particles
and, consequently, we have no propagating modes. So, we conclude that
the topological Chern-Simons term is compatible with the propagation
of the torsion as long as unitarity is to be respected.

We understand that the Chern-Simons term does not alter the conditions
for unitarity due to its lower derivative character and by virtue
of its topological aspect. Actually, as we know, a Chern-Simons term
alone does not yield local perturbations that we may identify as particles.
The same should not be true for other parity-breaking terms (as the
ones listed in Section \ref{sec:Description-of-The}), since they
carry derivatives of a higher order. We have no concrete arguments
in favor of this possibility, but , to our mind, they should be investigated.
Also, we expect no modification in the particle content of the massive
sector since, by analyzing the possible degrees of freedom that can
be propagated, all massive modes have been consistently excited. The
genuine massless spin-$2$ non-massive mode (corresponding to the
massless graviton) should not exist in three dimensions by a simple
counting of the on-shell degrees of freedom. It remains to be elucidated
if the remaining massless modes could propagate in a way compatible
with unitarity.

\begin{acknowledgments}
The authors are very grateful to Prof. A. J. Accioly for a critical
reading of our work. They also express their gratitude to CNPq-Brazil
and FAPERJ-Rio de Janeiro for the invaluable financial support.
\end{acknowledgments}
\appendix

\section*{Appendix: Inverse Matrices}

The inverse matrices that appears in the propagators are given by:

\begin{equation}
a^{-1}\left(0\right)=\frac{1}{D_{0}}\left(\begin{array}{cccc}
A_{11}^{\left(0\right)} & A_{12}^{\left(0\right)} & A_{13}^{\left(0\right)} & A_{14}^{\left(0\right)}\\
A_{12}^{\left(0\right)} & A_{22}^{\left(0\right)} & A_{23}^{\left(0\right)} & A_{24}^{\left(0\right)}\\
A_{13}^{\left(0\right)} & A_{23}^{\left(0\right)} & A_{33}^{\left(0\right)} & A_{34}^{\left(0\right)}\\
-A_{14}^{\left(0\right)} & -A_{24}^{\left(0\right)} & -A_{23}^{\left(0\right)} & A_{44}^{\left(0\right)}\end{array}\right),\label{eq:InvSpin0}\end{equation}
where \begin{eqnarray}
 & D_{0} & =8p^{2}\left[\left(u+r\right)\left(\left(u+2r+\left(\beta-\gamma\right)p^{2}\right)\left(\left(\left(u-t-s\right)\xi p^{2}-4t\left(u-s\right)\right)\right)-64d^{2}p^{2}\left(u-t-s\right)\right),\right.\nonumber \\
 &  & \left.-2r^{2}\left(\left(u-t-s\right)\xi p^{2}-4t\left(u-s\right)\right)\right],\label{eq:den0}\\
 & A_{11}^{\left(0\right)} & =4(u+r)(-4t(u-s)+(u-t-s)\lyxmathsym{\textgreek{x}}p^{2}))p^{2},\nonumber \\
 & A_{12}^{\left(0\right)} & =-2\sqrt{2}r\left(2\left(u-t-s\right)\left(-4t+\xi p^{2}\right)-8t^{2}\right)p^{2},\nonumber \\
 & A_{22}^{\left(0\right)} & =\left(\left(2u+4r+2\left(\beta-\gamma\right)p^{2}\right)\left(2\left(u-t-s\right)\left(-4t+\xi p^{2}\right)-8t^{2}\right)-256d^{2}p^{2}\left(u-t-s\right)\right)p^{2},\nonumber \\
 & A_{13}^{\left(0\right)} & =64dt\left(r+u\right)p^{2},\nonumber \\
 & A_{14}^{\left(0\right)} & =-32\sqrt{2}id\left(u-t-s\right)\left(r+u\right)\sqrt{p^{2}}p^{2},\nonumber \\
 & A_{22}^{\left(0\right)} & =4((u+2r+(\lyxmathsym{\textgreek{b}}-\lyxmathsym{\textgreek{g}})p^{2})(-4(u-s)t+(u-t-s)\lyxmathsym{\textgreek{x}}p\text{\texttwosuperior})-64d^{2}p^{2}(u-t-s))p^{2},\nonumber \\
 & A_{23}^{\left(0\right)} & =-64\sqrt{2}drtp^{2},A_{24}^{\left(0\right)}=64idr\left(u-t-s\right)\sqrt{p^{2}}p^{2},\nonumber \\
 & A_{33}^{\left(0\right)} & =4(-4t+\lyxmathsym{\textgreek{x}}p^{2})(u(3r+u)+(u+r)(\lyxmathsym{\textgreek{b}}-\lyxmathsym{\textgreek{g}})p^{2})-256p^{2}d^{2}(u+r),\nonumber \\
 & A_{34}^{\left(0\right)} & =-8\sqrt{2}i\sqrt{p^{2}}t\left(u(3r+u)+(\lyxmathsym{\textgreek{b}}-\lyxmathsym{\textgreek{g}})(u+r)p^{2}\right),\nonumber \\
 & A_{44}^{\left(0\right)} & =8p^{2}\left(u-t-s\right)\left(u(3r+u)+(\lyxmathsym{\textgreek{b}}-\lyxmathsym{\textgreek{g}})(u+r)p^{2}\right).\nonumber \end{eqnarray}

\begin{equation}
a^{-1}\left(1\right)=\frac{1}{D_{1}}\left(\begin{array}{cccc}
A_{11}^{\left(1\right)} & A_{12}^{\left(1\right)} & A_{13}^{\left(1\right)} & A_{14}^{\left(1\right)}\\
-A_{12}^{\left(1\right)} & A_{11}^{\left(1\right)} & -A_{14}^{\left(1\right)} & A_{13}^{\left(1\right)}\\
A_{13}^{\left(1\right)} & A_{14}^{\left(1\right)} & A_{33}^{\left(1\right)} & A_{34}^{\left(1\right)}\\
-A_{14}^{\left(1\right)} & A_{13}^{\left(1\right)} & -A_{34}^{\left(1\right)} & A_{33}^{\left(1\right)}\end{array}\right),\label{eq:InvSpin1}\end{equation}
where 

\begin{eqnarray}
 & D_{1} & =\frac{1}{2}p^{2}\left(4d^{2}\left(u-t\right)^{2}p^{2}-\left(\frac{1}{2}p^{2}\left(u-t\right)\beta-ut\right)^{2}\right),\label{eq:den1}\\
 & A_{11}^{\left(1\right)} & =-\frac{1}{4}\left(u-t\right)\left(\frac{1}{2}\left(u-t\right)\beta p^{2}-ut\right)p^{2},\nonumber \\
 & A_{12}^{\left(1\right)} & =-\frac{i}{2}\left(u-t\right)^{2}d\sqrt{p^{2}}p^{2},\nonumber \\
 & A_{13}^{\left(1\right)} & =-du\left(u-t\right)p^{2},\nonumber \\
 & A_{14}^{\left(1\right)} & =-\frac{i}{2}\sqrt{p^{2}}u\left(\frac{1}{2}\left(u-t\right)\beta p^{2}-ut\right),\nonumber \\
 & A_{33}^{\left(1\right)} & =-\left(\left(\frac{1}{2}p^{2}\beta-t\right)\left(u-t\right)^{2}+\left(\frac{1}{4}p^{4}\beta^{2}-2t^{2}\right)\left(u-t\right)-t^{2}\left(\frac{1}{2}\beta p^{2}+t\right)-4\left(u-t\right)d^{2}p^{2}\right),\nonumber \\
 & A_{34}^{\left(1\right)} & =-2idu^{2}\sqrt{p^{2}},\nonumber \end{eqnarray}

\begin{equation}
a^{-1}\left(2\right)=\frac{1}{D_{2}}\left(\begin{array}{cccc}
A_{11}^{\left(2\right)} & A_{12}^{\left(2\right)} & A_{13}^{\left(2\right)} & A_{14}^{\left(2\right)}\\
-A_{12}^{\left(2\right)} & A_{11}^{\left(2\right)} & -A_{14}^{\left(2\right)} & A_{13}^{\left(2\right)}\\
A_{13}^{\left(2\right)} & A_{14}^{\left(2\right)} & A_{33}^{\left(2\right)} & A_{34}^{\left(2\right)}\\
-A_{14}^{\left(2\right)} & A_{13}^{\left(2\right)} & -A_{34}^{\left(2\right)} & A_{33}^{\left(2\right)}\end{array}\right),\label{eq:InvSpin2}\end{equation}
where

\begin{eqnarray}
 & D_{2} & =2p^{2}\left(16d^{2}s^{2}p^{2}-\left(u\left(s-u\right)+p^{2}s\left(\beta+\gamma\right)\right)^{2}\right),\label{eq:den2}\\
 & A_{11}^{\left(2\right)} & =-s\left(su+s\left(\beta+\gamma\right)p^{2}-u^{2}\right)p^{2},\nonumber \\
 & A_{12}^{\left(2\right)} & =4ids^{2}\sqrt{p^{2}}p^{2},\nonumber \\
 & A_{13}^{\left(2\right)} & =-4dusp^{2},\nonumber \\
 & A_{14}^{\left(2\right)} & =iu\sqrt{p^{2}}\left(u\left(s-u\right)+s\left(\beta+\gamma\right)p^{2}\right),\nonumber \\
 & A_{33}^{\left(2\right)} & =-\frac{1}{s}\left(\left(u\left(s-u\right)+s\left(\beta+\gamma\right)p^{2}\right)^{2}-16s^{2}d^{2}p^{2}+u^{2}\left(u\left(s-u\right)+s\left(\beta+\gamma\right)p^{2}\right)\right),\nonumber \\
 & A_{34}^{\left(2\right)} & =-4idu^{2}\sqrt{p^{2}}.\nonumber \end{eqnarray}

\end{document}